\documentclass[10pt,a4paper, DIV=14]{scrartcl}
\usepackage[utf8]{inputenc}
\usepackage[T1]{fontenc}
\usepackage[english]{babel}
\usepackage{textcomp}
\usepackage{amsmath,amssymb}
\usepackage{lmodern}
\usepackage{graphicx}
\usepackage[dvipsnames,svgnames]{xcolor}
\usepackage{microtype}
\usepackage{hyperref} \hypersetup{pdfstartview=XYZ}

\usepackage{hyperref}

\usepackage{caption}
\usepackage{multirow}
\usepackage{supertabular}

\usepackage{tikz}
\usetikzlibrary{patterns}

\usepackage{authblk}

\makeatletter
\newcounter {subsubsubsection}[subsubsection]
\renewcommand\thesubsubsubsection{\thesubsubsection .\@alph\c@subsubsubsection}
\newcommand\subsubsubsection{\@startsection{subsubsubsection}{4}{\z@}%
                                     {-3.25ex\@plus -1ex \@minus -.2ex}%
                                     {1.5ex \@plus .2ex}%
                                     {\normalfont\normalsize\bfseries}}
\newcommand*\l@subsubsubsection{\@dottedtocline{3}{10.0em}{4.1em}}
\newcommand*{\subsubsubsectionmark}[1]{}
\makeatother

\title{Atomic energy levels population kinetics in dense plasmas - plasma ions electric microfield mixing dynamics effect (I): the general theory}
\author{Y.J. Aouad\footnote{PhD in Plasma Physics from UPMC "Université Pierre et Marie Curie" in the laboratory LULI "Laboratoire pour l'Utilisation des Lasers Intenses" - Ecole Polytechnique. Email: aouad852000@yahoo.fr}}
\affil{\footnotesize Theoretical Physics, Atomic Physics in Plasmas, France}

\begin{document}

\maketitle

\begin{abstract}

In the present paper we introduce a full and closed system of atomic kinetic equations on fractions of atomic energy levels $\{P_{\gamma J}\}$ taking into account the effect of the plasma ions electric microfield mixing dynamics. The system of equations that we present is deduced from a full quantum multi-atomic states $\{\mid \gamma J M >\}$ density matrix master equation. The analytical derivation of this atomic kinetic system of equations is based on the introduction of a new hierarchy chain of equations involving a new set of statistical atomic variables. The final system of atomic kinetic equations is obtained by introducing \textbf{relation de fermeture} to truncate the hierarchy chain of equations. We show that the final system of equations contains a new plasma ions electric microfield dependent atomic rates related to the process of mixing dynamics of atomic energy states. The analytical expression of this rate is given in this paper. We discuss the importance of the present analysis in view of the radiation emission originating from dense plasmas created by the interaction of the X-ray Free Electron Laser (XFEL’s) with solid density matter.
  
\end{abstract}

\tableofcontents
\bigskip

%%%%%%%%%%%%%%%%%%%%%%%%%%%%%%%%%%%%%%%%%%%%%%%%%%%%%%%%%%%%%%
\section{Introduction}

\subsection*{XFEL’s interaction with a solid state matter: creation of warm dense matter and strongly coupled plasmas}
\hspace{2mm} In the past few years, X-ray free electron lasers (XFEL's) (installations: LCLS 2011, XFEL 2011, SACLA XFEL 2011)  offer for the first time to the scientific community the opportunity to study new regimes of matter \cite{Rosmej4, Rosmej1, Youcef5}. The XFEL's has an outstanding properties: $10^{12}$ photons per pulse, photon energy in the XUV or X-ray range, short pulse duration of order of $10-100$ fs, high repetition frequency ($10-120$ Hz). These parameters has initiated an interest in different area of research: planetary science, astrophysics, inertial-confinement fusion, high-energy density physics, and the study of exotic states of matter never created in laboratories so far. 
\\
\\
Because the critical density of the XFEL’s is larger than the solid density, the absorption of the radiation energy of the laser is homogeneous and proceeds into the volume of the solid. This occurs on a short time scale when the matter is in its solid-state density allowing isochoric heating of the solid. The energy and intensity (of order of $10^{+16}$ W/cm$^{2}$) of the XFEL's may induces the intern shell photoionization of most atoms of the crystal and because the pulse duration of the XFEL's is of the order of the autoionization rate, hollow ions are formed and the solid can be transformed into an exotic state of matter: a hollow crystal \cite{Rosmej1, Galtier1}. The mechanisms of evolution of the hollow crystal to warm dense matter to strongly coupled plasma are not well understood.     

\subsection*{X-ray emission originating from autoionizing atomic states and hollow ion configurations: time-resolved emission}
\hspace{2mm} Nevertheless the spectroscopic methods give some way for the study of dense plasma regimes. This is based on the properties of the XFEL’s itself, namely the creation of multi-excited states and hollow ion configurations by direct photoionization of the K and L atomic shells, as the energy per photon that constitutes the XFEL’s allows the ionization of the internal shells of atoms (ex: K and L) \cite{Rosmej5}. These configurations are created principally in the dense plasma regime during the laser heating of matter. The autoionizing configurations (and their corresponding spectral emission) are characterized by a very short lifetimes of about $1-10$ fs due to high autoionization rates ($\Gamma \approx 10^{+13} - 10^{+16}$ s$^{-1}$). The short lifetime leads likewise to an intrinsic time resolution for the corresponding dielectronic satellite emission. The high intensity of the XFEL’s laser causes high number of photoionization events of the internal atomic shells and high population of multi-excited states and hollow ion configurations. That is why the development of new spectroscopic methods based on the multi-excited and hollow ions atomic configurations is enormously advantageous for the study of the coupled plasma regimes. 
\\
\\
In dense plasma regimes,  the formalism of atomic density matrix was proposed \cite{Youcef5, Youcef6} for the calculation of the atomic spectroscopic properties of autoionizing configurations in dense plasmas, i.e. warm dense matter and strongly coupled plasmas. This is motivated by the importance of taking into account the effect of plasma ions electric microfield mixing dynamics of atomic energy states in the calculation of the atomic populations of autoionizing configurations as well as the spectral line shapes associated to the radiation originating from these configurations in the frame of the standard methods. This is due to the fact that for autoionizing configurations, electron densities justifying the statistical approach exceed solid density bringing serious doubts to the statistical approach even in dense plasmas. Because autoionizing configurations are characterized by a high autoionization rates \cite{Rosmej1}. Therefore, local thermodynamical equilibrium populations can only be achieved for electron densities above solid density and non local thermodynamical equilibrium populations have to be considered for almost all parameters of practical interest. But, at high densities the effect of the plasma ions electric microfield mixing dynamics has to be taken into account in the calculation of the atomic populations by the use of the density matrix kinetics model instead of the standard collisional-radiative one.   
\\
\\
However in numerical calculations for many levels system, the rank of the system of equations on matrix elements of the density matrix operator is very large and even more complicated than the standard collisionnal-radiative model that is already prohibitive for numerical calculations \cite{Youcef5}. For the density matrix formalism, the difficulty arises firstly from the fact that the set of atomic quantum states $\{\mid \gamma J M >\}$ is used instead of atomic energy levels $\{\gamma J\}$ and secondly from the nature of the density matrix formalism that introduces coherences (non-diagonal matrix elements) in addition to populations (diagonal matrix elements). The situation is more complicated when the autoionizing atomic configurations are considered in the calculation. These configurations leads to a huge number of atomic energy levels of different $\gamma J$ terms as they have an open atomic sub-shells and various coupling schemes of the momentums of the bound electrons have to be implemented. 
\\
\\
Therefor; in order to perform real simulations taking into account the plasma ions electric microfield mixing dynamics effect on autoionizing configurations, a full and closed system of equations on fractions $\{P_{\gamma J}\}$ of atomic energy levels only has to be derived. In the present paper, we present the general theory that leads to this system of equations. The final result is independent from all non-diagonal matrix elements and it includes only usual fractions of atomic energy levels. A new plasma ions electric microfield dependent atomic rate is introduced and it is related to the process of mixing dynamics of atomic energy states.

%%%%%%%%%%%%%%%%%%%%%%%%%%%%%%%%%%%%%%%%%%%%%%%%%%%%%%%%%%%%%%

\section{Plasma ions electric microfield dependent atomic kinetics: density matrix kinetic equation in dense plasmas}
\hspace{2mm} The effect of the plasma on the atomic kinetics concerns both, the non dissipative Hamiltonian  mixing of the atomic levels by the quasi-static ion electric microfield and the dissipative time evolution part induced by the multiple atomic processes in the plasma \cite{Youcef5, Youcef6}. The later part is often introduced in the frame of the collisional-radiative kinetics model where in the density matrix case the difference is given by the relaxation of the non-diagonal density matrix elements also called coherences. By denoting the set  $\{\mid \gamma J M > \}$ of unperturbed atomic states ($J$ total kinetic moment, $M$ the quantized z-component of the total kinetic moment and $\gamma$ additional quantum numbers) that are eigenstates of the unperturbed emitting ion Hamiltonian $\hat{H}_{0}$ corresponding to the set $\{E_{\gamma J M}\}$ of energy eigenvalues, such that:
\\
\begin{equation}\label{H0_alpha}
\hat{H}_{0} \mid \gamma J M > = E_{\gamma J M}  \mid \gamma J M >
\end{equation}
\\ 
The matrix elements representation of the density operator $\rho$ for the two unperturbed atomic states $\mid \gamma J M >$ and $\mid \gamma' J' M' >$ writes:
\\
\begin{equation}\label{rho_jk}
\rho^{\gamma J,\  \gamma' J'}_{M,\  M'} = < \gamma J M \mid \rho \mid \gamma ' J' M' > 
\end{equation}
\\ 
The time evolution master equation of $\rho^{\gamma J,\  \gamma' J'}_{M,\  M'}$ is given by \cite{Youcef5}:
\\
\begin{multline}\label{rho_tev_matrix_elements_2}
\frac{\partial \rho^{\gamma J,\  \gamma' J'}_{M,\  M'}}{\partial t} = -\frac{i}{\hbar} (E_{\gamma J M }-E_{\gamma' J' M' } )\rho^{\gamma J,\  \gamma' J'}_{M,\  M'}- \\
 \frac{i}{\hbar} \sum_{\gamma''J'', M''}\left(\hat{V}(\vec{E})^{\gamma J,\  \gamma'' J''}_{M,\  M''} \times \rho^{\gamma'' J'',\  \gamma' J'}_{M'',\  M'}-\rho^{\gamma J,\  \gamma'' J''}_{M,\  M''} \times \hat{V}(\vec{E})^{\gamma'' J'',\  \gamma' J'}_{M'',\  M'}\right)- \\
\frac{1}{2}\left(\sum_{\gamma''J'', M''}W^{\gamma J,\  \gamma'' J''}_{M,\  M''} +\sum_{\gamma''J'', M''}W^{\gamma' J',\  \gamma'' J''}_{M',\  M''} \right) \rho^{\gamma J,\  \gamma' J'}_{M,\  M'}+\\
\delta_{\gamma,\gamma'} \delta_{J,J'} \delta_{M,M'} \sum_{\gamma''J'', M''}W^{\gamma'' J'',\  \gamma J}_{M'',\  M} \times \rho^{\gamma'' J'',\  \gamma'' J''}_{M'',\  M''}
\end{multline}
\\
where $\hbar$ is the reduced Planck constant and the first term of the right hand side of Eq.\ref{rho_tev_matrix_elements_2} includes  $\hat{H}_{0}$ the unperturbed Hamiltonian of the emitting ion Eq.\ref{H0_alpha}, $\hat{V}$ the interaction of the emitting ion with the static electric microfield $\vec{E}$ of the surrounding ions in the dipole approximation. More precisely, the use of the static approximation is justified by the fact that the relaxation time of the density matrix elements is smaller than the inverse plasma frequency associated to the ions. Denoting the dipole atomic moment of the emitting ion $\hat{\vec{d}}$, the interaction potential energy $\hat{V}$ is given by: 
\\
\begin{equation}\label{V_op}
\hat{V}(\vec{E})=-\hat{\vec{d}} \cdot \vec{E}
\end{equation}
\\
The second term in right hand-side of Eq.\ref{rho_tev_matrix_elements_2} contains the matrix elements of the interaction operator $\hat{V}(\vec{E})$ Eq.\ref{V_op}. By representing the electric field in the z-direction axis and using the Wigner-Eckart theorem, the non-diagonal matrix elements of $\hat{V}(\vec{E})$ write:
\\
\begin{equation}\label{ME_H_elec}
\left\langle \gamma J M \mid \hat{V}(\vec{E}) \mid\gamma ' J' M' \right\rangle = -e \ a_0 \ E \ (-1)^{J-M} \left(\begin{array}{clcr}
J & 1 & J'\\
-M & 0 & M'  \end{array}\right) \times < \gamma J \mid\mid \textbf{\emph{P}}^{\textbf{(1)}} \mid\mid \gamma ' J' > 
\end{equation}
\\
where $e$ is the electric charge of the electron, $a_0$ is the Bohr radius, $E$ is the amplitude of the plasma ions electric microfield, the expression in brackets $(...)$ is the 3-$j$ symbol and $< \gamma J \mid\mid \textbf{\emph{P}}^{\textbf{(1)}} \mid\mid \gamma ' J' >$ is the reduced matrix elements of the dipole moment  $\textbf{\emph{P}}^{\textbf{(1)}}$ of the emitter in units of $ea_0$. The properties of the 3-$j$ symbol lead to the selection rule $M = M'$ ($J$, $1$ and $J'$  obey the triangle relations for non-vanishing 3-$j$ symbol, $\delta(J1J')=+1$). The dissipative part in Eq.\ref{rho_tev_matrix_elements_2} is expressed in terms of the standard transition rates $W^{\gamma J,\ \gamma' J'}_{M,\ M'}$ from the atomic state $\mid \gamma J M >$ to the atomic state $\mid \gamma' J' M' >$ and it involves all collisionnal and radiative relaxation processes that populate and depopulate elements of the density matrix, i.e., spontaneous emission rate $A$ and autoionization rate $\Gamma$, collisionnal relaxation between atomic states that are mixed by the electric microfield $\vec{E}$ of ions and so on:
 \\
\begin{equation}\label{sum_W_process}
W^{\gamma J,\ \gamma' J'}_{M,\ M'} = \sum_{P \in \{Processes\}} \left(W^{\gamma J,\ \gamma' J'}_{M,\ M'}\right)^{P}
\end{equation}
\\
It is to note that the solution density matrix of Eq.\ref{rho_tev_matrix_elements_2} is field dependent: 
\\
\begin{equation}\label{rho_E}
\rho \equiv \rho(\vec{E})
\end{equation}
\\
If one is interested by the calculation of the atomic population of the atomic state $\mid \gamma J M >$, the final result must be averaged over the stationary plasma ions electric microfield distribution function $Q(\vec{E})$:
\\
\begin{equation}\label{rho_average}
<\rho^{\gamma J,\  \gamma J}_{M,\ M}>=\int d\vec{E} \ Q(\vec{E}) \ \rho^{\gamma J,\  \gamma J}_{M,\ M}(\vec{E})
\end{equation}
\\
However, the calculation of the spectral lines shape uses the field dependent density matrix $\rho(\vec{E})$ before the field average procedure \cite{Youcef5}.

%%%%%%%%%%%%%%%%%%%%%%%%%%%%%%%%%%%%%%%%%%%%%%%%%%%%%%%%%%%%%%

\section{Atomic density matrix relationship to populations of atomic energy levels $\{\gamma J\}$}
\hspace{2mm} The use of the density matrix kinetics Eq.\ref{rho_tev_matrix_elements_2} leads to some difficulties. The first one is related to the increasing of the rank of the system of equations due to the presence of the non-diagonal density matrix elements (coherences). The second difficulty is related to the projection procedure into matrix elements that is based on atomic quantum states $\mid \gamma J M >$ instead of atomic energy levels $\gamma J$. The number of atomic energy levels to be introduced for plasma spectroscopy simulations is very huge and different methods was introduced to decrease this number by using configurations (C) and superconfigurations (SC) instead of detailed energy levels $\gamma J$ accounting. From this point of view, in order to use the density matrix system of equations Eq.\ref{rho_tev_matrix_elements_2} in plasma spectroscopy it is necessary to rewrite it in terms of equations on energy levels $\gamma J$ to reduce the system by removing the $M$ dependence. The procedure of reduction is based on summing equations of type Eq.\ref{rho_tev_matrix_elements_2}.  
\\
\\
In this section we relate diagonal density matrix elements to populations $N_{\gamma J}$ of atomic energy levels $\gamma J$ by a sum procedure. From the physical point of view, this is motivated by the fact that for the calculation of spectroscopic observables, the atomic populations $N_{\gamma J}$ are used and the detail accounting of populations $N_{\gamma J M}$ of atomic quantum states is not needed, where: 
\\
\begin{equation}\label{N_GammaJ_GammaJM}
N_{\gamma J} =\sum_{-J \leq M \leq J} N_{\gamma J M} 
\end{equation}
\\
From a technical point of view, this is a first step for the reduction of the rank of the system of equations Eq.\ref{rho_tev_matrix_elements_2} leading to a more convenient system of equations for numerical calculations. Firstly, we discuss the limiting case where the plasma ions electric microfield vanishes such that $\vec{E}=\vec{0}$ in Eq.\ref{rho_tev_matrix_elements_2}  and then the general equations are given in the case $\vec{E} \neq \vec{0}$.

%%%%%%%%%%%%%%%%%%%%%%%%%%%%%%%%%%%%%%%%%%%%%%%%%%%%%%%%%%%%%%

\subsection{Absence of the plasma ions electric microfield mixing: case where $\vec{E}=\vec{0}$}
\hspace{2mm} In the absence of the plasma ions electric microfield mixing, the atomic density matrix system of equations in the atomic quantum states basis $\{\mid \gamma J M>\}$ writes:
\\
\begin{multline}\label{rho_tev_matrix_E0}
\frac{\partial \rho^{\gamma J,\  \gamma' J'}_{M,\  M'}}{\partial t} = -\frac{i}{\hbar} (E_{\gamma J M }-E_{\gamma' J' M' } )\rho^{\gamma J,\  \gamma' J'}_{M,\  M'} - \\
\frac{1}{2}\left(\sum_{\gamma''J'', M''}W^{\gamma J,\  \gamma'' J''}_{M,\  M''} +\sum_{\gamma''J'', M''}W^{\gamma' J',\  \gamma'' J''}_{M',\  M''} \right) \rho^{\gamma J,\  \gamma' J'}_{M,\  M'}+\\
\delta_{\gamma J,\gamma'J'} \delta_{M,M'} \sum_{\gamma''J'', M''}W^{\gamma'' J'',\  \gamma J}_{M'',\  M} \times \rho^{\gamma'' J'',\  \gamma'' J''}_{M'',\  M''}
\end{multline}
\\ 
It is to note that the energies $E_{\gamma J M}$ are $M$ independent by the degeneracy of atomic quantum states:
\\
\begin{equation}\label{E_gamaJM_J}
E_{\gamma J M} = E_{\gamma J} \ ; \  -J \leq M \leq J
\end{equation}
\\
Thus, from Eq.\ref{rho_tev_matrix_E0}, the diagonal matrix elements of the atomic density matrix are solutions of a pure standard collisional-radiative model:
\\
\begin{equation}\label{rho_tev_matrix_E0_Pop}
\frac{\partial \rho^{\gamma J,\  \gamma J}_{M,\  M}}{\partial t} = -  \rho^{\gamma J,\  \gamma J}_{M,\  M} \times \sum_{\gamma''J'', M''}W^{\gamma J,\  \gamma'' J''}_{M,\  M''}+\sum_{\gamma''J'', M''}W^{\gamma'' J'',\  \gamma J}_{M'',\  M} \times \rho^{\gamma'' J'',\  \gamma'' J''}_{M'',\  M''}
\end{equation}
\\ 
and the time dependent solution of the non-diagonal density matrix elements (coherences, $\gamma J M \neq \gamma' J' M'$) is given by:
\\
\begin{equation}\label{ATM_Coherences_Solution}
\rho^{\gamma J,\  \gamma' J'}_{M,\  M'}(t) = \rho^{\gamma J,\  \gamma' J'}_{M,\  M'}(0) \ \times \ e^{-\frac{i}{\hbar} (E_{\gamma J}-E_{\gamma' J'} ) \times t } \ \times \ e^{-\frac{1}{2} \left(\sum_{\gamma'',J'', M''}W^{\gamma J,\  \gamma'' J''}_{M,\  M''} +\sum_{\gamma'',J'', M''}W^{\gamma' J',\  \gamma'' J''}_{M',\  M''} \right) \times t }
\end{equation}
\\ 
which, from Eq.\ref{ATM_Coherences_Solution}, tends to zero in the permanent regime (the stationary solution):
\\
\begin{equation}\label{ATM_Coherences_Solution_Stationary}
\lim_{t \rightarrow +\infty} \rho^{\gamma J,\  \gamma' J'}_{M,\  M'}(t) = 0 
\end{equation}

%%%%%%%%%%%%%%%%%%%%%%%%%%%%%%%%%%%%%%%%%%%%%%%%%%%%%%%%%%%%%%

\subsubsection{Thermodynamic equilibrium limit}
\hspace{2mm} In the thermodynamic equilibrium conditions, the stationary solution of the atomic density matrix kinetic equations reads:
\\
\begin{equation}\label{Eq_DM_1}
\rho_{eq}=\frac{e^{-\beta \hat{H}_{0}}}{\text{Tr}(e^{-\beta \hat{H}_{0}})}
\end{equation}
\\
where $\beta=1/(k_B T_e)$, $k_B$ is the Boltzmann constant and $T_e$ is the electronic temperature. This solution Eq.\ref{Eq_DM_1} is purely diagonal in the unperturbed atomic quantum states basis representation $\mid \gamma J M >$:
\\
\begin{equation}\label{Eq_DM_alpha_beta}
\left[\rho_{eq}\right]^{\gamma J,\  \gamma' J'}_{M,\  M'} = \delta_{\gamma J,\gamma' J'} \delta_{M,M'}  \ \frac {e^{- \beta E_{\gamma J} }}{\sum_{\gamma'' J''}g_{\gamma'' J''} \times e^{ -\beta E_{\gamma'' J''}}}
\end{equation}
\\
where $g_{\gamma J}$ is the statistical weight associated to the atomic energy level $\gamma J$:
\\
\begin{equation}\label{Stat_weight_J}
g_{\gamma J} = 2J+1
\end{equation}
\\
Eq.\ref{Eq_DM_alpha_beta} leads to an interpretation of the diagonal elements of the atomic density matrix as the fractions $\{P_{\gamma J M} \}$ of the atomic populations associated to the set of atomic quantum states $\{\mid \gamma J M > \}$. By denoting $N_{tot}$ the total number of ions, $P_{\gamma J M}$ is given by:
\\
\begin{equation}\label{P_unperturbed_J_M}
P_{\gamma J M} \equiv \frac{N_{\gamma J M}}{N_{tot}} = \rho^{\gamma J,\  \gamma J}_{M,\  M}
\end{equation}
\\
and, from Eqs.\ref{N_GammaJ_GammaJM}, \ref{P_unperturbed_J_M} one deduces the fraction $P_{\gamma J}$ associated to the population of the atomic energy level $\gamma J$:
\\
\begin{equation}\label{P_unperturbed_J}
P_{\gamma J} \equiv \frac{N_{\gamma J}}{N_{tot}} = \sum_{-J \leq M \leq J} \rho^{\gamma J,\  \gamma J}_{M,\  M}
\end{equation}
\\
which leads, in the thermodynamic equilibrium limit Eqs.\ref{Eq_DM_alpha_beta}, \ref{P_unperturbed_J}, to: 
\\
\begin{equation}\label{P_Eq_unperturbed_J}
\left[P_{eq}\right]_{\gamma J} =  \frac {g_{\gamma J} \times e^{- \beta E_{\gamma J} }}{\sum_{\gamma'' J''}g_{\gamma'' J''} \times e^{ -\beta E_{\gamma'' J''}}}
\end{equation}
\\
Hereafter equations on the usual atomic energy levels fractions $\{P_{\gamma J}\}$ are derived. The essential thinks is that in the present context the influence of the plasma ions electric microfield is considered as well as the standard collisional-radiative processes in the plasma. 

%%%%%%%%%%%%%%%%%%%%%%%%%%%%%%%%%%%%%%%%%%%%%%%%%%%%%%%%%%%%%%

\section{Reduction of the rank of the atomic density matrix system of equations: equations on atomic energy levels fractions $\{P_{\gamma J}\}$ in dense plasmas}
\hspace{2mm} From Eqs.\ref{rho_tev_matrix_elements_2},\ref{P_unperturbed_J}, the fraction $P_{\gamma J}$ of the atomic energy level $\gamma J$ obeys to the following equation:
\\
\begin{equation}\label{tev_Popo_gammaJ_sum}
\frac{\partial P_{\gamma J}}{\partial t} = \sum_{-J \leq M \leq J} \frac{\partial \rho^{\gamma J,\ \gamma J}_{M, \ M}}{\partial t}
\end{equation}
\\
and by using Eq.\ref{ME_H_elec}, Eq.\ref{tev_Popo_gammaJ_sum} leads to:
\\
\begin{multline}\label{tev_Pop_gammaJ_gammaJ}
\frac{\partial P_{\gamma J}}{\partial t} = \frac{i}{\hbar} \ e \ a_0 \ E \ \times \sum_{\gamma'', J''} < \gamma J \mid\mid \textbf{\emph{P}}^{\textbf{(1)}} \mid\mid \gamma'' J'' > \times P^{\textbf{1},\ 0}_{\gamma'' J'',\ \gamma J} - \\
\frac{i}{\hbar} \ e \ a_0 \ E \ \times \sum_{\gamma'', J''} < \gamma'' J'' \mid\mid \textbf{\emph{P}}^{\textbf{(1)}} \mid\mid \gamma J > \times P^{\textbf{1}, \ 0}_{\gamma J,\ \gamma'' J''} - \\
P_{\gamma J} \sum_{\gamma'',J''} W_{\gamma J,\ \gamma'' J''} + \sum_{\gamma'', J''} W_{\gamma'' J'',\ \gamma J} \times P_{\gamma'' J''}
\end{multline}
\\
In Eq.\ref{tev_Pop_gammaJ_gammaJ}, we have introduced a set atomic statistical variables:
\\
\begin{equation}\label{Pop_gammaJ_NDS}
P^{\textbf{1}, \ 0}_{\gamma'' J'',\ \gamma J} \equiv \sum_{M, M''} (-1)^{J-M} \left(\begin{array}{clcr}
J'' & \ \ J & 1\\
M'' & -M & 0  \end{array}\right) \times \rho^{\gamma'' J'',\  \gamma J}_{M'',\  M} 
\end{equation}
\\
and the transition rate $W_{\gamma J,\ \gamma'' J''}$ from the atomic energy level $\gamma J$ to the atomic energy level $\gamma'' J''$ is given by:
\\ 
\begin{equation}\label{W_gammaJ_gammaJ}
W_{\gamma J,\ \gamma'' J''}= \sum_{M} \frac{\rho^{\gamma J,\  \gamma J}_{M,\  M}}{P_{\gamma J}} \sum_{M''} W^{\gamma J,\  \gamma'' J''}_{M,\  M''}
\end{equation}
\\
For practical calculations, the rates $W_{\gamma J,\ \gamma'' J''}$ Eq.\ref{W_gammaJ_gammaJ} are calculated by introducing the statistical weighted average in the sum:
\\ 
\begin{equation}\label{W_gammaJ_gammaJ_Stat}
W_{\gamma J,\ \gamma'' J''}=  \frac{1}{g_{\gamma J}} \sum_{M,\ M''} W^{\gamma J,\  \gamma'' J''}_{M,\  M''}
\end{equation}
\\
Eqs.\ref{tev_Pop_gammaJ_gammaJ},\ref{Pop_gammaJ_NDS} show that the atomic population fraction $P_{\gamma J}$ of the the energy level $\gamma J$ is not depending on the detailed non-diagonal elements accounting for the quantum number $M$. It depends only on a 3-$j$ weighted sum of them Eq.\ref{Pop_gammaJ_NDS}. In order to close the system of equations Eq.\ref{tev_Pop_gammaJ_gammaJ}, we introduce equations on variables $P^{\textbf{1},\ 0}_{\gamma J,\ \gamma' J'}$ Eq.\ref{Pop_gammaJ_NDS}. So, we proceed as follows:
\\
\begin{equation}\label{tev_P_gammaJ_gammaJ}
\frac{\partial P^{\textbf{1},\ 0}_{\gamma J,\ \gamma' J'}}{\partial t} = \sum_{M, M'} (-1)^{J'-M'} \left(\begin{array}{clcr}
J & \ \ J' & 1\\
M & -M' & 0  \end{array}\right) \times \frac{\partial \rho^{\gamma J,\  \gamma' J'}_{M,\  M'}}{\partial t} 
\end{equation}
\\
then from Eqs.\ref{rho_tev_matrix_elements_2},\ref{tev_P_gammaJ_gammaJ}, we get:
\\
\begin{multline}\label{tev_P_gammaJ_gammaJ_2}
\frac{\partial P^{\textbf{1},\ 0}_{\gamma J,\ \gamma' J'}}{\partial t} = -\frac{i}{\hbar} (E_{\gamma J}-E_{\gamma' J'} ) \times P^{\textbf{1},\ 0}_{\gamma J,\ \gamma' J'} + \\
\frac{i}{\hbar} \ e \ a_0 \ E \ \times < \gamma J \mid\mid \textbf{\emph{P}}^{\textbf{(1)}} \mid\mid \gamma' J' > \times \left(P^{\textbf{J11},\ 00}_{\gamma' J',\ \gamma' J'} - P^{\textbf{J'11}, \ 00}_{\gamma J,\ \gamma J}\right)+ \\
\frac{i}{\hbar} \ e \ a_0 \ E \ \times \sum_{\gamma'', J'' \neq \gamma', J'}  < \gamma J \mid\mid \textbf{\emph{P}}^{\textbf{(1)}} \mid\mid \gamma'' J'' > \times P^{\textbf{J11},\ 00}_{\gamma'' J'',\ \gamma' J'} - \\
\frac{i}{\hbar} \ e \ a_0 \ E \ \times \sum_{\gamma'', J'' \neq \gamma, J} < \gamma'' J'' \mid\mid \textbf{\emph{P}}^{\textbf{(1)}} \mid\mid \gamma' J' > \times P^{\textbf{J'11}, \ 00}_{\gamma J,\ \gamma'' J''} -\\
\frac{1}{2}\left( \sum_{\gamma'',J''} W_{\gamma J,\ \gamma'' J''} +  \sum_{\gamma'',J''} W_{\gamma' J',\ \gamma'' J''} \right) \times P^{\textbf{1},\ 0}_{\gamma J,\ \gamma' J'}
\end{multline}
\\
where, we have introduced a second set of atomic statistical variables $P^{\textbf{J11},\ 00}_{\gamma'' J'',\ \gamma' J'}$ given by:
\\
\begin{equation}\label{P1100_gammaJ_gammaJ}
P^{\textbf{J11},\ 00}_{\gamma'' J'',\ \gamma' J'} = \sum_{M,\ M',\ M''} (-1)^{J-M} (-1)^{J'-M'} \left(\begin{array}{clcr}
J & \ \ J' & 1\\
M & -M' & 0  \end{array}\right) \times \left(\begin{array}{clcr}
J & \ \ J'' & 1\\
M & -M'' & 0  \end{array}\right) \times \rho^{\gamma'' J'',\  \gamma' J'}_{M'',\  M'}
\end{equation}
\\
The collisional-radiative relaxation terms $W_{\gamma J,\ \gamma'' J''}$ of $P^{\textbf{1},\ 0}_{\gamma J,\ \gamma' J'}$ in Eq.\ref{tev_P_gammaJ_gammaJ_2} are calculated as in Eq.\ref{W_gammaJ_gammaJ_Stat}. From Eq.\ref{tev_P_gammaJ_gammaJ_2}, it is shown that the sum Eq.\ref{tev_P_gammaJ_gammaJ} leads to the introduction of new variables $P^{\textbf{J11},\ 00}_{\gamma'' J'',\ \gamma' J'}$ Eq.\ref{P1100_gammaJ_gammaJ} which involves a multiplication of two 3-$j$ ($\sum (\text{3-}j) \times (\text{3-}j)$) symbols instead of one as in the definition of variables $P^{\textbf{1},\ 0}_{\gamma J,\ \gamma' J'}$ Eq.\ref{Pop_gammaJ_NDS} ($\sum (\text{3-}j)$). The introduction of equations on variables $P^{\textbf{J11},\ 00}_{\gamma'' J'',\ \gamma' J'}$ will involve new variables that are linear combinations of terms involving an increasing number of 3-$j$ symbols in their respectives factors ($\sum (\text{3-}j) \times (\text{3-}j) \times (\text{3-}j)$). This is a hierarchy chain of equations on variables that are series expansions involving more and more 3-$j$ symbols in coefficients. Hereafter in section 5 \ref{Truncation} we give a formula to close the system of equations by introducing a procedure allowing the truncation of the hierarchy.

%%%%%%%%%%%%%%%%%%%%%%%%%%%%%%%%%%%%%%%%%%%%%%%%%%%%%%%%%%%%%%

\subsection{Physical interpretation of the new variables $P^{\textbf{1},\ 0}_{\gamma J,\ \gamma' J'}$}
\hspace{2mm} In the following we show that the new variables $P^{\textbf{1},\ 0}_{\gamma' J',\ \gamma J}$ are related to one of components of the so-called state multipole or statistical tensor of rank $1$: $\left<T(\gamma J, \gamma' J')^{\dag}_{1Q}\right>$ and $-1 \leq Q \leq +1$. The component in question corresponds to $Q = 0$. 
\\
\\
So, by decomposing the density operator $\rho$ in terms of the irreducible tensor operators $T(\gamma' J', \gamma J)_{KQ}$ where $\vert J'-J \vert \leq K \leq (J'+J)$ and  $-K \leq Q \leq +K$: 
\\
\begin{equation}\label{rho_tenors_set}
\rho = \sum_{\gamma'J',\ \gamma J,\ K,\ Q} \left<T(\gamma' J', \gamma J)^{\dag}_{KQ}\right> T(\gamma' J', \gamma J)_{KQ}
\end{equation}
\\
in Eq.\ref{rho_tenors_set}, the irreducible tensor operators $T(\gamma' J', \gamma J)_{KQ}$ are defined in terms of 3-$j$ symbols by:
\\
\begin{equation}\label{irreducible_tenors_set}
T(\gamma' J', \gamma J)_{KQ} = \sum_{M',\ M} (-1)^{J'-M'} \sqrt{2K+1} \left(\begin{array}{clcr}
J' & \ \ J & K\\
M' & -M & -Q  \end{array}\right) \mid \gamma' J' M' ><\gamma J M \mid
\end{equation}
\\
and the state multipoles or statistical tensors are given by:
\\
\begin{equation}\label{state_multipoles_set}
 \left<T(\gamma' J', \gamma J)^{\dag}_{KQ}\right> = \sum_{M',\ M} (-1)^{J'-M'} \sqrt{2K+1} \left(\begin{array}{clcr}
J' & \ \ J & K\\
M' & -M & -Q  \end{array}\right) \rho^{\gamma' J',\ \gamma J}_{M',\ M}
\end{equation}
\\
Then from Eqs.\ref{Pop_gammaJ_NDS},\ref{state_multipoles_set}, we find that the new variables $P^{\textbf{1},\ 0}_{\gamma J,\ \gamma' J'}$ are related to $\left<T(\gamma J, \gamma' J')^{\dag}_{10}\right>$ by:
\\
\begin{equation}\label{Relation_P_Stat_tensors}
P^{\textbf{1},\ 0}_{\gamma J,\ \gamma' J'} =  \frac{1}{3} \left<T(\gamma J, \gamma' J')^{\dag}_{10}\right>^{*}
\end{equation}
\\
From a physical point of view $P^{\textbf{1},\ 0}_{\gamma J,\ \gamma J}$ is related to the component $\left<T(\gamma J)^{\dag}_{10}\right>$ (equivalently: $\left<T(\gamma J, \gamma J)^{\dag}_{10}\right>$) of the so-called orientation vector $\left<T(\gamma J)^{\dag}_{1Q}\right>$ $(Q=0,\pm 1)$ (equivalently: $\left<T(\gamma J, \gamma J)^{\dag}_{1Q}\right>$). More precisely,  $P^{\textbf{1},\ 0}_{\gamma J,\ \gamma J}$ is related to the expectation value $\left<\hat{J}_{z}\right>$ of the z-component $\hat{J}_{z}$ of the total kinetic moments operator $\hat{\vec{J}}$:
\\
\begin{equation}\label{Relation_P_Jz}
P^{\textbf{1},\ 0}_{\gamma J,\ \gamma J} =  \left(\frac{1}{3(2J+1)(J+1)J}\right)^{\frac{1}{2}} \times \left<\hat{J}_{z}\right>
\end{equation}

%%%%%%%%%%%%%%%%%%%%%%%%%%%%%%%%%%%%%%%%%%%%%%%%%%%%%%%%%%%%%%

\section{General discussion for numerical applications in real simulations of atomic kinetics in dense plasmas}
\hspace{2mm} At this step we have introduced a hierarchy chain of equations. Starting with equations on fractions of atomic populations $\{P_{\gamma J}\}$ Eq.\ref{P_unperturbed_J}, a new set of variables is introduced $\{P^{\textbf{1},\ 0}_{\gamma J,\ \gamma' J'}\}$ Eq.\ref{Pop_gammaJ_NDS} and so on and so forth. But in practice for numerical calculations, one has to truncate the hierarchy chain of equations. In the following, we truncate the hierarchy chain of equations at the level of the set of variables $\{P^{\textbf{J11},\ 00}_{\gamma J,\ \gamma' J'}\}$ Eq.\ref{P1100_gammaJ_gammaJ}.

\subsection{Truncation of the hierarchy chain of equations: evaluation of variables $P^{\textbf{J11},\ 00}_{\gamma J,\ \gamma' J'}$}\label{Truncation}
\hspace{2mm} The analysis of equation Eq.\ref{tev_P_gammaJ_gammaJ_2} on variables $P^{\textbf{1},\ 0}_{\gamma J,\ \gamma' J'}$ shows that the variables $P^{\textbf{J11},\ 00}_{\gamma J,\ \gamma' J'}$ enter this equation in two different contributions. The first is into the second term in Eq.\ref{tev_P_gammaJ_gammaJ_2} which is proportional to the difference $(P^{\textbf{J11},\ 00}_{\gamma' J',\ \gamma' J'}-P^{\textbf{J'11},\ 00}_{\gamma J,\ \gamma J})$ which involves only diagonal matrix elements of the atomic density matrix in the basis of the set of atomic states $\{\mid \gamma J M >\}$ due to the selection rules imposed by the 3-$j$ symbols in Eq.\ref{P1100_gammaJ_gammaJ}. We call this contribution: \textbf{first kind contribution}. And secondly into the third and the forth terms in Eq.\ref{tev_P_gammaJ_gammaJ_2} that are sums of terms proportionals to $P^{\textbf{J11},\ 00}_{\gamma'' J'',\ \gamma' J'}$ (where, $\gamma'' J'' \neq \gamma J$) and $P^{\textbf{J'11},\ 00}_{\gamma J,\ \gamma'' J''}$ (where, $\gamma'' J'' \neq \gamma' J'$), these terms involve only non-diagonal elements of the atomic density matrix in the basis of the set of atomic states $\{\mid \gamma J M >\}$ that are indirect on the evolution of $P^{\textbf{1},\ 0}_{\gamma J,\ \gamma' J'}$ considered as a connection $\gamma J \leftrightarrow \gamma' J'$. We call this contribution: \textbf{second kind contribution}.  
\\
\\
To truncate the hierarchy chain of equations, in Eq.\ref{tev_P_gammaJ_gammaJ_2} we take into account only the \textbf{first kind contribution} and we neglect the \textbf{second kind contribution}. The \textbf{first kind contribution} is expressed in terms of fractions of atomic energy levels $\{P_{\gamma J}\}$ and is evaluated as follows:
\\
\begin{equation}\label{first_kind}
P^{\textbf{J11},\ 00}_{\gamma' J',\ \gamma' J'}-P^{\textbf{J'11},\ 00}_{\gamma J,\ \gamma J} = C^{\textbf{J11},\ 00}_{\gamma' J'} \times \frac{P_{\gamma' J'}}{g_{\gamma' J'}}  - C^{\textbf{J'11},\ 00}_{\gamma J} \times \frac{P_{\gamma J}}{g_{\gamma J}}
\end{equation}
\\
where in Eq.\ref{first_kind}, $g_{\gamma J}$ is the statistical weight of the energy level $\gamma J$ and the coefficients $C^{\textbf{J11},\ 00}_{\gamma' J'}$ are purely numerical values expressed in terms of sums involving 3-$j$ symbols and are given by:
\\
\begin{equation}\label{C1100_gammaJ}
C^{\textbf{J11},\ 00}_{\gamma' J'} = \sum_{M,\ M',\ M''} (-1)^{J-M} (-1)^{J'-M'} \left(\begin{array}{clcr}
J & \ \ J' & 1\\
M & -M' & 0  \end{array}\right) \times \left(\begin{array}{clcr}
J & \ \ J' & 1\\
M & -M'' & 0  \end{array}\right)
\end{equation}
\\
The general idea of this truncation procedure in the evaluation of the set of variables $\{P^{\textbf{J11},\ 00}_{\gamma J,\ \gamma' J'}\}$ is motivated by a thermodynamic-like \textbf{relation de fermeture} of the distribution of atomic states $\gamma J M$ inside the atomic energy level $\gamma J$. As in Eq.\ref{Eq_DM_1}, the density operator in the thermodynamic limit is purely diagonal so that we considered a statistical distribution of $\gamma J M$ inside the atomic energy level $\gamma J$ in the evaluation of the \textbf{first kind contribution} and a vanishing \textbf{second kind contribution} as it involves only indirect non-diagonal elements of the atomic density matrix (indirect regarding to the $\gamma J \leftrightarrow \gamma' J'$ connection).
\\
\\
Thus, under this  \textbf{relation de fermeture} the equation Eq.\ref{tev_P_gammaJ_gammaJ_2} on the variable $P^{\textbf{1},\ 0}_{\gamma J,\ \gamma' J'}$ Eq.\ref{Pop_gammaJ_NDS} and taking into account Eq.\ref{first_kind} leads to:
\\
\begin{multline}\label{tev_P_gammaJ_second_kind}
\frac{\partial P^{\textbf{1},\ 0}_{\gamma J,\ \gamma' J'}}{\partial t} = -\frac{i}{\hbar} (E_{\gamma J}-E_{\gamma' J'} ) \times P^{\textbf{1},\ 0}_{\gamma J,\ \gamma' J'} + \\
\frac{i}{\hbar} \ e \ a_0 \ E \ \times < \gamma J \mid\mid \textbf{\emph{P}}^{\textbf{(1)}} \mid\mid \gamma' J' > \times \left(C^{\textbf{J11},\ 00}_{\gamma' J'} \times \frac{P_{\gamma' J'}}{g_{\gamma' J'}}  - C^{\textbf{J'11},\ 00}_{\gamma J} \times \frac{P_{\gamma J}}{g_{\gamma J}}\right) -\\
\frac{1}{2}\left( \sum_{\gamma'',J''} W_{\gamma J,\ \gamma'' J''} +  \sum_{\gamma'',J''} W_{\gamma' J',\ \gamma'' J''} \right) \times P^{\textbf{1},\ 0}_{\gamma J,\ \gamma' J'}
\end{multline}
\\
Thus a closed system of equations is obtained by combining Eq.\ref{tev_Pop_gammaJ_gammaJ} and Eq.\ref{tev_P_gammaJ_second_kind} coupling the set of variables $\{P_{\gamma J}\textbf{;} \ P^{\textbf{1},\ 0}_{\gamma J,\ \gamma' J'}\}$. In the next section, we show that a stationary version of this closed system of equations leads to a system of equations involving only fractions of atomic energy levels $\{P_{\gamma J}\}$. This reduce considerably the rank of the initial system and is very practical for numerical simulations that take into account the dynamics of the plasma ions electric microfield mixing effect.

%%%%%%%%%%%%%%%%%%%%%%%%%%%%%%%%%%%%%%%%%%%%%%%%%%%%%%%%%%%%%%
\section{Stationary system of kinetic equations on fractions of atomic energy levels $\{P_{\gamma J}\}$ in dense plasmas: plasma ions electric microfield mixing dynamics effect}
\hspace{2mm} The stationary solution of Eq.\ref{tev_P_gammaJ_second_kind} leads to an analytical expression of the set of variables $\{P^{\textbf{1},\ 0}_{\gamma J,\ \gamma' J'}\}$ in terms of the set of atomic fractions $\{P_{\gamma J}\}$ of atomic energy levels. This analytical solution writes:
\\
\begin{equation}\label{Analytical_Solution_P_gammaJ_second_kind}
P^{\textbf{1},\ 0}_{\gamma J,\ \gamma' J'}= T_{\gamma J,\ \gamma' J'} \times \left(C^{\textbf{J11},\ 00}_{\gamma' J'} \times\frac{P_{\gamma' J'}}{g_{\gamma' J'}}  - C^{\textbf{J'11},\ 00}_{\gamma J} \times \frac{P_{\gamma J}}{g_{\gamma J}}\right)
\end{equation}
\\
where in Eq.\ref{Analytical_Solution_P_gammaJ_second_kind}, the term $T_{\gamma J, \gamma' J'}$ is given by:
\\
\begin{equation}\label{T_gammaJ_gammaJ}
T_{\gamma J,\ \gamma' J'} = \frac{\left[\frac{i}{\hbar} \ e \ a_0 \ E \ \times < \gamma J \mid\mid \textbf{\emph{P}}^{\textbf{(1)}} \mid\mid \gamma' J' >\right]}{\left[\frac{i}{\hbar} (E_{\gamma J}-E_{\gamma' J'} ) +\frac{1}{2}\left( \sum_{\gamma'',J''} W_{\gamma J,\ \gamma'' J''} +  \sum_{\gamma'',J''} W_{\gamma' J',\ \gamma'' J''} \right)\right] }
\end{equation}
\\
By inserting the analytical solution Eq.\ref{Analytical_Solution_P_gammaJ_second_kind} into the stationary version of the system of equations on fractions of atomic populations $\{P_{\gamma J}\}$ of atomic energy levels $\{\gamma J\}$, we get:
\\
\begin{multline}\label{Stat_Pop_gammaJ_gammaJ}
0 = -P_{\gamma J} \times \sum_{\gamma'', J''} \left[  W_{\gamma J,\ \gamma'' J''} + W(\gamma J , \  \gamma'' J'', E) \times \frac{C^{\textbf{J''11},\ 00}_{\gamma J}}{g_{\gamma J}} \right] + \\
\sum_{\gamma'', J''} P_{\gamma'' J''} \times \left[ W_{\gamma'' J'',\ \gamma J} + W(\gamma J , \  \gamma'' J'', E) \times \frac{C^{\textbf{J11},\ 00}_{\gamma'' J''}}{g_{\gamma'' J''}}\right]
\end{multline}
\\
where in Eq.\ref{Stat_Pop_gammaJ_gammaJ}, the term $W(\gamma J , \  \gamma'' J'', E)$ is given by:
\\
\begin{equation}\label{W_E_gammaJ_gammaJ}
W(\gamma J , \  \gamma'' J'', E) =- \left( S_{\gamma'' J'' , \  \gamma J}  \times T_{\gamma J,\ \gamma'' J''}  + S_{\gamma J , \  \gamma'' J''}  \times T_{\gamma'' J'',\ \gamma J}  \right) 
\end{equation}
\\
where in Eq.\ref{W_E_gammaJ_gammaJ}, the term $S_{\gamma J , \  \gamma'' J''}$ is given by:
\\
\begin{equation}\label{S_gammaJ_gammaJ}
S_{\gamma J , \  \gamma'' J''} = \frac{i}{\hbar} \ e \ a_0 \ E \ \times < \gamma J \mid\mid \textbf{\emph{P}}^{\textbf{(1)}} \mid\mid \gamma'' J'' > 
\end{equation}
\\
The explicit formula for the rate $W(\gamma J , \  \gamma'' J'', E)$ is then given by Eqs.\ref{T_gammaJ_gammaJ},\ref{W_E_gammaJ_gammaJ},\ref{S_gammaJ_gammaJ}:
\\
\begin{multline}\label{W_E_gammaJ_gammaJ_2}
W(\gamma J , \  \gamma'' J'', E) = \left(\frac{e a_{0} E}{\hbar}\right)^{2} \times < \gamma J \mid\mid \textbf{\emph{P}}^{\textbf{(1)}} \mid\mid \gamma'' J'' > \times < \gamma'' J'' \mid\mid \textbf{\emph{P}}^{\textbf{(1)}} \mid\mid \gamma J > \times \\
\frac{\left[ \sum_{\gamma'''J'''} W_{\gamma J,\ \gamma''' J'''} +  \sum_{\gamma'''J'''} W_{\gamma'' J'',\ \gamma'''J'''} \right]}{\left[ \frac{E_{\gamma J}-E_{\gamma'' J''}}{\hbar} \right]^2 +\frac{1}{4}\left[ \sum_{\gamma'''J'''} W_{\gamma J,\ \gamma''' J'''} +  \sum_{\gamma'''J'''} W_{\gamma'' J'',\ \gamma''' J'''} \right]^2}
\end{multline}
%{\left[ \frac{E_{\gamma J}-E_{\gamma' J'}}{\habr} \right]^2 +\frac{1}{4}\left[ \sum_{\gamma'',J''} W_{\gamma J,\ \gamma'' J''} +  \sum_{\gamma'',J''} W_{\gamma' J',\ \gamma'' J''} \right]^2}
\\
The system of equations Eq.\ref{Stat_Pop_gammaJ_gammaJ} is a full and closed system of equations on fractions $\{P_{\gamma J}\}$ of atomic energy levels ${\gamma J}$. This system of equations take into account a new rates that introduce the effect of the plasma ions electric microfield mixing dynamics of atomic energy levels. 
%%%%%%%%%%%%%%%%%%%%%%%%%%%%%%%%%%%%%%%%%%%%%%%%%%%%%%%%%%%%%%

\section{Summary}
\hspace{2mm} In this paper we have proposed a full and closed system of atomic kinetic equations on fractions of atomic energy levels $\{P_{\gamma J}\}$ taking into account the effect of the plasma ions electric microfield mixing dynamics. This system of equations contains a new plasma ions electric microfield dependent atomic rates related to the process of mixing dynamics of atomic energy states. The analytical expression of this rate is given in this paper. We discussed the importance of the present analysis in view of the radiation emission originating from dense plasmas created by the interaction of the X-ray Free Electron Laser (XFEL’s) with solid density matter. This new system of atomic kinetic equations could allow studying the influence of the correlations between ions on the atomic populations in the warm dense matter (WDM) and the strongly coupled plasmas (DSCP). 
     
%%%%%%%%%%%%%%%%%%%%%%%%%%%%%%%%%%%%%%%%%%%%%%%%%%%%%%%%%%%%%%

%%%%%%%%%%%%%%%%%%%%%%%%%%%%%%%%%%%%%%%%%%%%%%%%%%%%%%%%%%%%%%


\begin{thebibliography}{0}

\bibitem[Rosmej, 2007]{Rosmej4}
F.B. Rosmej, R. W. Lee, Europhysics Letters \textbf{77}, 24001 (2007).

\bibitem[Rosmej, 2012]{Rosmej1}
F.B. Rosmej, \textit{Exotic states of High density matter driven by intense XUV/X-ray Free Electron Lasers}, Free Electron Laser, InTech 2012, editor S. Varró, p. 187 – 212, ISBN 978-953-51-0279-3.

\bibitem[Aouad et al., 2017]{Youcef5}
Y.J. Aouad, F.B. Rosmej, A.V. Demura, V.S. Lisitsa,  \textit{Atomic population kinetics in the context of XUV/X-ray free electron laser generating warm dense matter and strongly coupled plasmas}, arxiv:1702.00680v1, (2017).

\bibitem[Aouad, 2017]{Youcef6}
Y.J. Aouad,  \textit{Stark profiles modeling of radiation lines originating from atomic autoionizing states in dense plasmas, solid state matter under short intense XUV/X-ray free electron laser irradiation}, arxiv:1702.02940v1, (2017).

\bibitem[Galtier et al., 2011]{Galtier1}
E. Galtier, F. B. Rosmej, T. Dzelzainis, D. Riley, F. Y. Khattak, P. Heimann, R.W. Lee, A. J. Nelson, S. M. Vinko, T. Witcher, B. Nagler, J. S. Wark, T. Tschenscher, S. Toleikis, R. Fäustlin, R. Sobierajski, M. Jurek, L. Juha, J. Chalupsky, V. Hajkova, M. Kozlova, and J. Krzywinski. \textit{Decay of crystalline order and equilibration during solid-to plasma transition induced by 20 fs microfocused 92 ev free electron laser pulses}, Phys. Rev. Lett \textbf{106}, 164801 (2011).

\bibitem[Rosmej, 1997]{Rosmej5}
F.B. Rosmej, J. Phys. B Lett.: At. Mol. Opt. Phys. \textbf{30}, L819 (1997).

\end{thebibliography}
\end{document}